# The citation-based indicator and combined impact indicator ― New options for measuring impact[*]

Ping Zhou    Yongfeng Zhong

Institute of Scientific and Technical Information of China

No. 15 Fuxing Road, Beijing 100038, China.

E-Mail: zhoup@istic.ac.cn; zhongyf@istic.ac.cn

**Abstract**

Metrics based on percentile ranks (PRs) for measuring scholarly impact involves complex treatment because of various defects such as overvaluing or devaluing an object caused by percentile ranking schemes, ignoring precise citation variation among those ranked next to each other, and inconsistency caused by additional papers or citations. These defects are especially obvious in a small-sized dataset. To avoid the complicated treatment of *PRs* based metrics, we propose two new indicators – the citation-based indicator (*CBI*) and the combined impact indicator (*CII*). Document types of publications are taken into account. With the two indicators, one would no more be bothered by complex issues encountered by *PRs* based indicators. For a small-sized dataset with less than 100 papers, special calculation is no more needed. The *CBI* is based solely on citation counts and the *CII* measures the integrate contributions of publications and citations. Both virtual and empirical data are used so as to compare the effect of related indicators. The *CII* and the PRs based indicator *I3* are highly correlated but the former reflects citation impact more and the latter relates more to publications.

## 1. Introduction

In the bibliometric community, many methods and measures have been proposed and developed for measuring scholarly impact, among which are Garfield's Impact Factor (*IF*) (Garfield, 1972; Garfield & Sher, 1963) and Leiden's indicator sets (e.g., *Journal Citation Rank, Field Citation Rank,* and *Citations per Publications*) (Moed et al., 1995, Van Raan, 2006). All these indicators are measured based on averaging methods and have been applied in performance evaluation. Nevertheless, in recent years such kind of metrics has been challenged and alternatives have been proposed (e.g., Adams et al., 2007; Lundberg, 2007; Opthof & Leydesdorff, 2010 ; Bormann, 2010; van Raan, et al., 2010; Waltman, et al, (2010); Gingras & Larivière, 2011; Leydesdorff & Bornmann, 2011).

---

[*] This research was supported by the National Natural Science Foundation of China (NSFC) with grant number 71073153.

The key argument of the challengers is that citation distribution can be highly skewed and any measure of central tendency is theoretically meaningless. In this context, Bornmann and Mutz (2011) proposed to classify citation distributions into six percentile ranks (6PR) which are top 1%, to-5%, top-10%, top-25%, top-50%, and bottom-50%. Application of percentile ranks makes it possible to compare distributions of citations across unequally sized document sets using a single scheme for the evaluation of the shape of the distribution. This approach was extended to hundred percentile by others (e.g., Leydesdorff, et al., 2011).

But the methodology of Bornmann and Mutz (2011) still involves average – they averaged over the percentile ranks. Based on the percentile rank method, Leydesdorff & Bornmann (2011) proposed a new measurement – the *Integrated Impact Indicator* (*I3*), which applies percentiles to rank documents according to their citation counts and integrates the rank and frequencies of the rank. The *I3* is formulated as follows:

$$I3 = \sum_i x_i * f(x_i) \tag{1}$$

Where $x_i$ is the frequency of papers in each percentile, and $f(x_i)$ is the percentile rank of each paper. The indicator has at least the following two advantages: Firstly, it takes both the size and the shape of the citation distribution into account. Secondly, it can be applied to any units of analysis (e.g., journals, nations, universities, institutions, individuals). The *I3* measurement has achieved recognition, for instance, Rousseau considers *I3* a congruous indicator of absolute performance (2011, 2012).

However, both percentile rank and the *I3* have shortcomings, because 1) They give the maximum rank to those with the same citation counts, which may overvalue an object with long tail of citation distribution or with many equal citation counts; 2) They ignore the precise citation variation among those at different rank positions, which may de-value those in a higher position but with significantly higher citation counts than the one ranked next; 3) Inconsistency may occur when additional papers or citations are taken into consideration (Schreiber, 2012).

To make it clearer, we create an extreme case with a set of publications and corresponding citations as shown in Table 1. To avoid complexity of calculating different document types of publications, which will be discussed later, let us suppose all the publications here belong to one document type (e.g., either article or review or proceeding paper).

Several ways can be used for calculating percentile ranks. Here we just discuss two of them – "[(b+e)/n] * 100" and "[b/n] * 100". Where b is the number of scores below x, e is the number of scores equal to x, and n is the number of scores. With formula "[(b+e)/n] ∗ 100", uncited publications E, F, G, H, I, and J would be ranked the 6th, and publications B, C, and D will be ranked the 2nd with percentile of 90%. Although paper A receives citations much more than the rest and contributes 92.5% to the citations of the whole set, it is only 1 position higher than the one in the 2nd position. The one position variation between papers A and B cannot precisely reflect the excellent citation performance of publication A. In other words, the citation impact of publication A is severely devalued and the weight to the rest is over-valued. Noticing the overvaluation of uncited publications Leydesdorff & Bornmann (2012) has revised the calculation of *I3* in their Letter to the Editor of JASIST by applying the formula "[b/n] ∗ 100".

**Table 1.** Impact measured with different metrics

| Publications | A | B | C | D | E | F | G | H | I | J | Total |
|---|---|---|---|---|---|---|---|---|---|---|---|
| **Citations** | 111 | 3 | 3 | 3 | 0 | 0 | 0 | 0 | 0 | 0 | 120 |
| **Percentile Ranks [(b+e)/n]*100** | 100 | 90 | 90 | 90 | 60 | 60 | 60 | 60 | 60 | 60 | 730 |
| **Percentile Ranks [b/n]*100** | 90 | 60 | 60 | 60 | 0 | 0 | 0 | 0 | 0 | 0 | 270 |
| *CBI* | 92.5 | 2.5 | 2.5 | 2.5 | 0 | 0 | 0 | 0 | 0 | 0 | 100 |

With formula "[b/n] ∗ 100", over-valuing effect of uncited publications can be prevented. But devaluing/over-valuing effect still exists, and impact of uncited publications cannot be measured. Take publications A and B in Table 1 for example, percentile of publications A is only 30% higher than that of publication B, while citation counts of the former is 36 times higher than the latter.

To solve the problem of devaluing/over-valuing effect of metrics based on percentile ranks in the case of less than 100 papers in the reference set, Schreiber (2012) suggested a fractional scoring rule. But "the fractional scoring makes the determination of the weights rather complicated in the general case" (p12, Schreiber, 2012). The other problem of inconsistency of metrics based on percentile ranks has also been solved by Schreiber, but the non-linearity of the weights for the different percentile ranks can still lead to changes in the ranking (Schreiber, 2012).

By far, a strong impression may emerge: Metrics based on percentile ranks are rather complicated. Too many issues have to be considered in assigning ranks, which include selection of ranking schemes (i.e., "[(b+e)/n] ∗ 100" or "[b/n] ∗ 100" or others), inconsistency and other problems of dataset with less than 100 paper, application of Schreiber's fractional scoring method so as to realize exact evaluation, and so on. To avoid the complexity of the mentioned metrics, we proposed alternatives that are much simpler and can be applied regardless of the size of a dataset. One solution is to solely focus on citations and is labeled as *Citation-Based Indicator* (*CBI*). The other takes both publications and citations into account and is thus named as *Combined Impact Indicator* (*CII*). Both indicators measure impact from different perspectives. Comparison between the new indicators and those based on percentile ranks will also be processed in an empirical case.

## 2. The Citation-Based Indicator (CBI)

When only citations are considered for measuring impact, the citation-based indicator (*CBI*) can be applied. It is a consensus that being cited is an indication of peer recognition, although many reasons may cause citation behavior (Glänzel, 2008, p59). Highly cited publications are considered as having high impact. The following formula illustrates how the *CBI* is calculated:

$$CBI = 100 \times \sum r_d \sum_{i=1}^{n} \frac{c_i}{c_d} \qquad (2)$$

Where $C_i$ denotes citations received by paper $i$ of one document type (e.g., either article, or review, or proceeding papers). The $C_d$ represents total citations received by publications of this document type. The $r_d$ is the ratio of the number of publications of the current document type in the total publication set. With the $r_d$ variation of citation chances of different publication types can be normalized within a reference set. Without using the $r_d$ for normalization, document types of publications with low ratio in a whole publication set may get benefit in measuring impact.

To better illustrate how document types affect the results of the *CBI*, we create an example illustrating *CBI* value before and after being normalized with $r_d$ (Table 2). In total, both journals A and B have published the same number of papers (i.e., 135) and received the same citation counts (i.e., 240), and thus both have the same value of citation per paper (1.778 = 240/135). When document types of publications are considered and simply add fraction of each document type together, citation impact values of the two journals vary significantly: Un-normalized the *CBI* value of Journal B (1.867) would be much higher than that of Journal A (1.133).

**Table 2.** Un-normalized and normalized journal *CBI*.

|   |              | *Article* | *Review* | *Proceeding paper* | *Total* | *c/p* | *CBI(1)* | *CBI(2)* |
|---|--------------|-----------|----------|--------------------|---------|-------|----------|----------|
| A | Publications | 120       | 10       | 5                  | 135     | 1.778 | 113.3    | 54.6     |
|   | Citations    | 120       | 119      | 1                  | 240     |       |          |          |
| B | Publications | 80        | 50       | 5                  | 135     | 1.778 | 186.7    | 45.4     |
|   | Citations    | 80        | 150      | 10                 | 240     |       |          |          |

*CBI(1)* – Before being normalized with the ratio of publications of a document type ($r_d$) in a reference set.
*CBI(2)* – After being normalized with the ratio of publications of a document type ($r_d$) in a reference set.

Nevertheless, only taking document types into account regardless of different ratios of document types in a data set still has defect in measuring impact. In a dataset, the number of publications of different document types varies greatly. Articles usually take the largest proportion, while reviews, proceeding papers and letters take the least. Let us still use Table 2 as an example. Proceeding papers take the least proportion in the two journals but may significantly affect the *CBI* value.

Average citations received by review papers in journal A is 11.9 (= 119/10) and that of journal B is 3 (= 150/50). This huge difference, however, cannot be well reflected if simply add up contribution of review papers to the *CBI* [for journal A: 119/(119+150) = 0.442; journal B: 150/(119+150) = 0.558]. The same problem occurs in measuring contributions of proceeding papers to the *CBI*. The five proceeding papers in journal A only received one citation with an average citation per paper of 1/5, while the other five proceeding papers in journal B received 10 citations with an average citations per paper of 2 (= 10/5). The 10 proceeding citations of journal B affect the *CBI* value significantly (Table 3).

**Table 3.** Contribution of each document type to journal *CBI*.

| Journal | Article CBI | Review CBI | CBI of Proceeding Paper | Journal CBI |
|---|---|---|---|---|
| *A* | 60.0 | 44.2 | 9.1 | 113.3 |
| *B* | 40.0 | 55.8 | 90.9 | 186.7 |

When document types are considered in measuring citation impact, those receiving low citation counts would take advantage if citation ratio of a subset in each document type is simply added up. In a dataset, citation counts of a specific document type may be decided by two factors in addition to each paper's research topic of the document type. The first factor is the document type itself. For instance, review papers can be cited more than others. The second factor is the relative ratio of publications of a document type in a dataset. It is common that a journal publishes more articles than reviews, proceedings papers and letters. With formula 2, the $\sum_{i=1}^{n} \frac{c_i}{c_d}$ measures the impact of the first factor, and $r_d$ normalizes the influence of the second factor. In the later part of the paper, correlation between different indicators including the *CBI* (before and after normalization with $r_d$) will be discussed based on an empirical case of journals.

**3. The Combined Impact Indicator (CII)**

With the *CBI*, only cited publications can be measured. But uncited publications may also have their values (Ingwersen et al., 2000). As Glänzel (2008, p59) pointed out, "a paper uncited several years after publications gives information about its reception by colleagues but does not reveal anything about its quality or the standing of its author(s)". A comprehensive measurement for research performance should take both cited and uncited publications into account. The indicator *I3* (Leydesdorff & Bornmann, 2011) contributes greatly in this respect, although somewhat complex as mentioned in the introduction part. With the combined impact indicator (*CII*), however, exact citation counts (of cited publications) and uncited publications can be measured without being bothered by the complex issues involved in metrics based on percentile ranks. Formula 3 illustrates how the *CII* of a subset j (e.g., a journal, an organization, a country, etc.) in a reference set is measured:

$$CII = 100 * [\sum r_d \sum_{i=1}^{n} \frac{c_i}{C_d} + \sum(r_d * \frac{1}{C_d} * \frac{n_{dj}-n_{dju}}{n_{dj}} * n_{dju})] \qquad (n_{dj} > 0) \qquad (3)$$

$$= 100 * \sum r_d (\sum_{i=1}^{n} \frac{c_i}{C_d} + \frac{1}{C_d} * \frac{n_{dj}-n_{dju}}{n_{dj}} * n_{dju})$$

In formula 3, the first part measures the impact of cited publications (i.e., *CBI*) of a subset *j*. The second part measures impact of uncited publications, where $n_{dj}$ denotes total number of publications of a document type in subset *j*, and $n_{dju}$ represents the number of uncited publications of the document type in subset *j*. In the second part of formula 3, the $\frac{1}{C_d}$ is the average weight of one citation of publications in one document type. The $\frac{n_{dj}-n_{dju}}{n_{dj}}$ represents the ratio of cited publications of a document type in subset *j* ($n_{dj} > 0$). For $n_{dj} = 0$, the value of the second part in formula 3 would be 0.

In our opinion, measuring impact of uncited publications of a subset should consider three factors including the number of uncited publications, cited publication ratio and the average weight of one citation received by publications of a document type. The weight of uncited publications of a subset should be in accordance with ratio of cited publications of the subset: A subset with higher cited publication ratio should be endowed higher weight to its uncited publications. Thus impact of uncited publications can be measured by the expression "$\frac{n_{dj}-n_{dju}}{n_{dj}} * n_{dju}$". But impact weight of uncited publications should be lower than that of cited. This condition can be guaranteed by multiplying "$\frac{n_{dj}-n_{dju}}{n_{dj}} * n_{dju}$" with the average weight of one citation of publications in one document type (i.e., $\frac{1}{C_d}$). The function of $r_d$ is the same as that in formula 2.

## 4. Correlations between relevant indicators

To compare and illustrate the results of relevant indicators, we downloaded data for journals in the subject category of library science and information science from the *Journal Citation Report* (social sciences 2010 version) of Thomson Reuters. In total the *JCR* covered 77 journals in the subject category in 2008 and 2009. But two journals (i.e., the *Libraries & the Cultural Recordand* and the *Informacao & Sociedade-Estudos*) did not publish any papers in this period, leaving 75 journals for analysis. The 75 journals in all published 5788 citable documents (articles, proceeding papers and reviews) in the two years. In calculating *I3*, percentile ranks are based on formula "(b/n) ∗ 100" so as to avoid over-valuing effect of "[(b+e)/n] ∗ 100" on uncited publications. Articles, reviews and proceeding papers are included. Both Spearman's rank order correlations and Pearson correlations are analyzed (Table 4).

As expected, there exist high correlations between the total citations (*TC*) and the other indicators (i.e., the *CBI*, *CII*, and *I3*) because of size effect. But normalization with the ratio of publication types in a reference set can further improve correlations. For example, the Pearson *r* between the *CBI* and the *NP* and *TC* increases from 0.626 to 0.844 and from 0.883 to 0.999 respectively, and that between the *CII* and *NP* and *TC* rise from 0.643 to 0.861 and from 0.886 to 0.996. Significant increase of correlations between the *CII* and *I3* also happens with such normalization (Table 4). These facts imply that document types of publications should be considered in measuring impact, because different types of publications inherently vary in terms of chances of being cited. In latter correlation analysis, values of the *CII(2)* and *CBI(2)* will be used to represent the *CII* and the *CBI*.

**Table 4.** Pearson correlations *r* (upper triangle) and rank order correlations (Spearman's $\rho$; lower triangle) for various indicators based on 75 journals in library and information science.

|        | NP      | TC      | IF      | CBI(1)  | CBI(2)  | CII(1)  | CII(2)  | I3      |
|--------|---------|---------|---------|---------|---------|---------|---------|---------|
| **NP**     |         | .830**  | .290*   | .626**  | .844**  | .643**  | .861**  | .873**  |
| **TC**     | .585**  |         | .671**  | .883**  | .999**  | .886**  | .996**  | .984**  |
| **IF**     | .312**  | .936**  |         | .747**  | .646**  | .734**  | .627**  | .617**  |
| **CBI(1)** | .463**  | .924**  | .911**  |         | .859**  | .999**  | .848**  | .823**  |
| **CBI(2)** | .604**  | .997**  | .920**  | .902**  |         | .864**  | .999**  | .989**  |
| **CII(1)** | .471**  | .913**  | .895**  | .998**  | .891**  |         | .854**  | .833**  |
| **CII(2)** | .648**  | .990**  | .891**  | .884**  | .996**  | .874**  |         | .994**  |
| **I3**     | .648**  | .989**  | .892**  | .895**  | .993**  | .885**  | .994**  |         |

**p* = 0.01 (2-tailed). **p* = 0.05 (2-tailed).
*NP* – Number of Publications; *TC* – Times Cited; *IF2010* – Impact Factor in 2010.
*CBI(1)* – Before normalizing with the ratio of publications of a document type in a reference set.
*CBI(2)* – After normalizing with the ratio of publications of a document type in a reference set.
*CII(1)* - Before normalizing with the ratio of publications of a document type in a reference set.
*CII(2)* – After normalizing with the ratio of publications of a document type in a reference set.

Because of differentiating contributions of document types of publication, the *CBI* does not overlap with the *TC*. Strong correlations exist between the *CII* and *I3* (*r* = 0.994, $\rho$ = 0.995). Although both indicators significantly correlates with size (i.e., *TC* and *NP*), the *CII* correlates with *TC* better (r = 0.996, $\rho$ = 0.990) than the *I3* (r = 0.984, $\rho$ = 0.989), whereas the latter correlates with the number of publications (i.e., *NP*) (*r* = 0.873, $\rho$ = 0.648) better (*r* = 0.867, $\rho$ = 0.646). But such correlation difference is not significant.

To better illustrate attribution of the indicators in Table 4, we drew a plot of the (varimax-rotated) two-factor solution (Figure 1). These indicators are classified into three groups with the *IF* and the number of publications (*NP*) belonging to two distinct groups close to the horizontal and vertical coordinates respectively. The third group is composed of the *CII*, *I3*, *CBI* and *TC* standing between the *IF* and *NP* but closer to the *NP*. The three groups of indicators measure contribution of productivity (i.e., number of publications) and peer recognition (i.e., citation counts) to impact differently. The *IF* group measures average citation impact, the *NP* group represents productivity, and the group including the *CII*, *I3*, *CBI* and *TC* take both productivity and peer recognition into account.

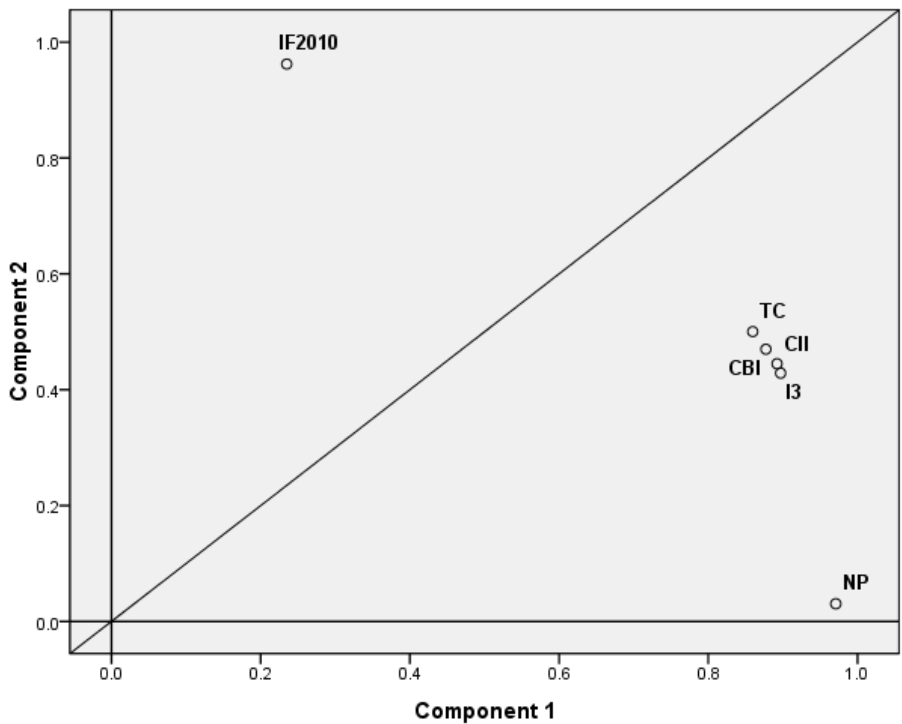

**Figure 1.** Varimax rotated two-factor solution of six variables.

When document types of publications are not considered, the *CBI* is another form of *TC* and the two indicators would overlap with each other completely in a plot like Figure 1. The distance between the *CBI* and *TC* in Figure 1 implies that document types do play a role in measuring citation impact. The *CII* and *I3* have similar attribution but the *CII* stands closer to citations while *I3* locates closer to publications.

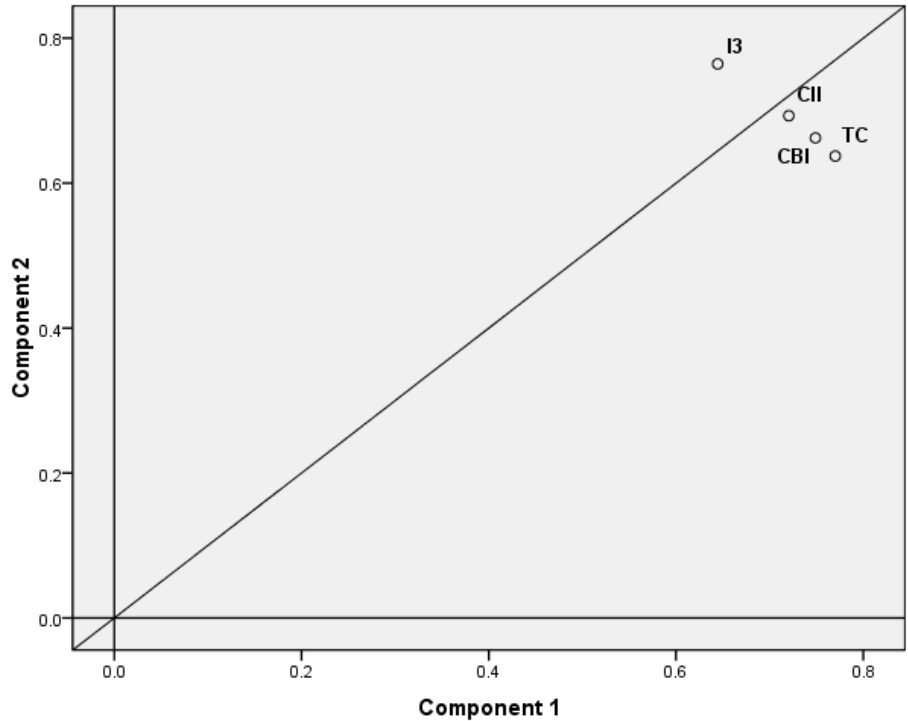

**Figure 2.** Varimax rotated two-factor solution of four variables.

In order to better differentiate attributions of the *CII*, *I3*, *CBI* and *TC*, we limited factor analysis for the four indicators and got the result shown in Figure 2. These indicators can still be classified: The *I3* represents one group and the other three (i.e., the *CII*, *CBI* and *TC*) form another one. Compared with *I3*, the *CII* respects more to citation counts and, thus, represents citation impact better. This conclusion is reasonable because the *CII* uses exact citation counts while the *I3* applies percentile ranks of citations. As mentioned before, percentile ranks may not well reflect exact difference of citation counts between those in different rank positions as illustrated in Table 1, even though in a large-sized dataset such a defect would not be significant (Schreiber, 2012).

## 5. Ranking journals with relevant indicators

To test the effect of different measurement, the 75 journals in Library and Information Science were ranked with related indicators. Table 5 lists the top-15 journals based on *CII* values. Rank positions using other indicators are marked respectively.

The ranking results with *CBI* and *CII* are most similar. In addition to screening the same top-four journals, rank difference of the rest journals is also not significant. Similarity in terms of ranking results between the *CII* and *I3* is also distinct, but rank difference of the rest journals is more obvious than that between *CBI* and *CII*. The largest rank difference is reflected on *Journal of Informetrics*: The journal is ranked 9th with the *CII* and 13th with the *I3*. With the *TC* one may also get similar result as those based on the *CII*, *CBI* and *I3*. But with *IF* the result would be amazingly different because of different perspectives: The former group correlates with size and the later emphasizes average effect.

The size or average effects is well reflected on the *JASIST* and the *MIS Quarterly*. With size advantage in both publications and citations, the *JASIST* has in total published 387 citable items in 2008 and 2009, and have received 845 citations in 2010. Whereas the *MIS Quarterly* only published 74 items that received 335 citations within the same time span. When size effect is considered, the *JAISIST* would be ranked higher than the *MIS Quarterly*. With the averaging effect of *IF*, however, the rank orders of the two journals would be reversed. Another journal that takes advantage of the average effect of the *IF* is the *Information Systems Research* (Table 5).

**Table 5.** Ranks of 15 journals in Library and Information Science subject category with the highest *CII* value in comparison with the number publications (*NP*), total citations (*TC*) and *IF* in 2010 and *I3*.

| Journals | TP | TC | IF2010 | CBI | CII | %I3 |
|---|---|---|---|---|---|---|
| *JASIST* | 387 | 845[1] | 2.183[8] | 12.88[1] | 14.17[1] | 10.57[1] |
| *Scientometrics* | 317 | 577[3] | 1.82[12] | 8.87[2] | 10[2] | 8.07[2] |
| *Journal of the American Medical Informatics Association* | 205 | 612[2] | 2.985[4] | 8.8[3] | 9.36[3] | 6.64[3] |
| *Information & Management* | 118 | 307[5] | 2.602[6] | 4.8[4] | 5.14[4] | 3.77[5] |
| *Information Processing & Management* | 165 | 280[6] | 1.697[13] | 4.38[6] | 4.91[5] | 4.57[4] |
| *MIS Quarterly* | 74 | 335[4] | 4.527[1] | 4.39[5] | 4.49[6] | 3.07[6] |
| *International Journal of Geographical Information Science* | 131 | 194[9] | 1.481[19] | 3[9] | 3.49[7] | 3.03[7] |
| *Journal of Management Information Systems* | 80 | 203[8] | 2.538[7] | 3.03[8] | 3.27[8] | 2.42[11] |
| *Journal of Informetrics* | 67 | 215[7] | 3.209[2] | 3.07[7] | 3.24[9] | 2.34[13] |
| *Government Information Quarterly* | 105 | 174[10] | 1.657[15] | 2.58[10] | 2.98[10] | 2.43[10] |
| *Journal of Health Communication* | 106 | 158[12] | 1.491[18] | 2.39[11] | 2.76[11] | 2.57[8] |
| *Journal of Computer-Mediated Communication* | 96 | 150[13] | 1.563[16] | 2.26[13] | 2.59[12] | 2.24[14] |
| *International Journal of Information Management* | 101 | 144[14] | 1.426[20] | 2.2[14] | 2.56[13] | 2.41[12] |
| *Journal of Information Science* | 101 | 143[15] | 1.416[21] | 2.13[15] | 2.48[14] | 2.44[9] |
| *Information Systems Research* | 53 | 170[11] | 3.208[3] | 2.36[12] | 2.44[15] | 1.99[16] |

## 6. Discussion and conclusions

The advantage of metrics based on percentile ranks is that citation distributions across unequally sized document sets can be compared by using a single scheme for evaluating the shape of the distribution. The indicator *I3* advances further by combining both the size and the shape of the distribution. However, metrics based on percentile rank may overvalue an object with many equally cited publications and may ignore the precise citation variation among those at different rank positions.

Citation counts are only used for ranking. Once a rank is determined, exact citation counts will not be used anymore in measuring impact regardless how great citation difference may exist between publications ranked next to each other (e.g., $1^{st}$ and $2^{nd}$ positions in Tables 1). Those in a higher position and have significantly higher citation counts than the one ranked next might be de-valued. Although excluding zero citations may significantly improve the results of *I3*, the over-valuing/devaluing effect cannot be eliminated completely. Schreiber's solution for exact evaluation of percentile ranks for a dataset with less than 100 papers is still complicated as he himself pointed out (Schreiber, 2012).

Different types of publications have different chances of being cited, thus measuring impact should consider such variation among document types. The ratio of publication types in a data set also plays a significant role in measuring impact. In the current paper, the effect of these factors has been considered in designing the indicators *CBI* and *CII*. Nonetheless, one should keep in mind that Thomson Reuters may misclassify document types of papers in the social sciences (Harzing, 2012). When only one document type is to be evaluated, the *CBI* would be of no difference from the commonly used indicator – percentage citations of a subset. But the *CII* still has its merit if impact of uncited publications is to be assessed.

In measuring academic impact, citations are still considered a good option, although shortcomings may exist (Westney, 1998). Gläzel and Schoepflin (1999) noted that citation was "one important form of use of scientific information within the framework of documented science communication". They consider citation as "a formalized account of the information use and can be taken as a strong indicator of reception at this level". Most cited publications receive recognitions from scholarly community even though negative comments may exist, and thus produce "impact". When only citations are considered in performance evaluation, the *CBI* can be used.

Regarding publication quantity, it is usually used to measure productivity and is rarely applied in impact evaluation. If a publication is not cited several years after its publication, "it is likely that the results involved do not contribute essentially to the contemporary scientific paradigm system of the subject field in question" (Braun et al., 1985). But being uncited within a period does not equal to uselessness. Uncited publications are also outcomes of researchers and should be considered in measuring comprehensive contributions with condition that their weight should be lower than those of cited. Taking the above factors into account, the *CII* may well serve the purpose of measuring integrated research impact.

The *CBI* can be applied when only citation impact is emphasized in evaluating research performance. When productivity is also considered, the *CII* is an option. Similar to the *I3*, the *CII* and *CBI* can be applied to evaluate any other objects such as universities, regions and countries within a reference set, although the current paper only used journals for empirical study. One may label both the *CII* and *I3* as indicators for measuring academic impact because of integrating values of productivity and citation counts. Although variation between the two exists, they are highly correlated, because both give weights to productivity and citations. Only the *CII* respects citation impact more than *I3* because of using exact citation counts in the measurement, while the *I3* correlates more with productivity.

The indicators *CII* and *CBI* can avoid the defects of metrics based on percentile ranks and application of the two indicators is simpler. As mentioned in the introduction part, indicators based on percentile ranks involve complex issues such as selection of ranking schemes, inconsistency and other problems of dataset with less than 100 paper, application of Schreiber's fractional scoring so as to exactly evaluate percentile ranks, and so on. With the *CII* and *CBI*, however, all the troublesome issues no more exist. The two new indicators can be applied to any datasets regardless of their size. The advantage of the two indicators is more obvious for a small dataset (e.g., individuals or university departments) – special treatment and inconsistency caused by citation changes of papers at different rank positions is unnecessary. Schreiber's fractional scoring is no more needed because of applying exact citations instead of percentile ranks in the calculation. Nevertheless, difference between the *I3* and the two new indicators (i.e., *CBI* and *CII*) would be marginal to a large dataset.

## Acknowledgements

This paper is originated from a thought of Ping Zhou. Not quite confident with her thought Ping Zhou communicated with Prof. Loet Leydesdorff from the University of Amsterdam and received valuable feedback. The contribution of Prof. Fred Y. Ye from Zhejiang University is also appreciated.